\documentclass[11pt,nohyper,a4paper]{JHEP3}
\usepackage{amssymb,euscript,bbold}
\newcommand{\sect}[1]{\setcounter{equation}{0}\section{#1}}

\newcommand{\bea}{\begin{eqnarray}}
\newcommand{\eea}{\end{eqnarray}}
\def\be{\begin{equation}}
\def\ee{\end{equation}}
\def\ba{\begin{eqnarray}}
\def\ea{\end{eqnarray}}
\def\nn{\nonumber \\}

\def\cL{{\cal{L}}}

\font\mybb=msbm10 at 10pt
\def\bb#1{\hbox{\mybb#1}}

\def\bR {\bb{R}}

\title{Uniqueness of Five-Dimensional Supersymmetric Black Holes}

\author{Jan B. Gutowski\\ Mathematical Institute\\
Oxford, OX1 3LB, UK\\ E-mail:
\email{gutowski@maths.ox.ac.uk}}

\abstract{A classification of supersymmetric
solutions of five dimensional ungauged supergravity coupled to arbitrary
many abelian vector multiplets is used to prove a uniqueness
 theorem for asymptotically flat supersymmetric black holes with regular
horizons. It is shown that the near-horizon geometries of solutions
for which the scalars and gauge field strengths are sufficiently
regular on the horizon are flat space, $AdS_3 \times S^2$,
or the near-horizon BMPV solution. Furthermore, the only black hole
which has the near-horizon BMPV geometry for its near-horizon
geometry is the solution found by Chamseddine and Sabra.}

\keywords{Supergravity Models, Black Holes in String Theory}

\preprint{}

\begin{document}

\setcounter{equation}{0}

\sect{Introduction}

In the past few years there has been a renewed interest in examining
various aspects of supergravity theories and their solutions.
Considerable progress has been made in our understanding of the
 structure of supersymmetric solutions of
several supergravity theories.
In particular, by generalising the methods originally used by Tod
to analyse certain four-dimensional supergravities \cite{toda}, \cite{todb}
it has been possible to construct classifications of supersymmetric
solutions of several  supergravities in four, five and six dimensions
\cite{classa}, \cite{classb}, \cite{classc}, \cite{classd} which provide a
comprehensive description of such solutions (although the 1/2 and
(possibly) 3/4 supersymmetric solutions which we expect to occur in the
minimal gauged five-dimensional supergravity \cite{classb} have yet to
be fully analysed).

Rather less is known about the structure of solutions in
higher dimensional supergravities. For example, although
all maximally supersymmetric solutions of eleven dimensional
supergravity are known \cite{max11}, and all solutions preserving
1/32 of supersymmetry have been classified \cite{gaunpak11},
\cite{gaungutpak11}; solutions preserving intermediate proportions
of supersymmetry have yet to be systematically classified.
Nevertheless, one can still use similar methods to examine
certain restricted classes of higher dimensional solutions;
some recent examples of such analysis are \cite{class2}
and \cite{class3}. All of this activity has produced a very large number
of new supergravity solutions, many of which have interesting properties.
It is clear that considerable work remains to be done in this area.

One particularly useful application of the lower dimensional
classifications is in black hole physics. This is of some
interest, because of the important role which string theory has played
in our understanding of the microscopic origin of the entropy
of certain four and  five-dimensional black holes \cite{stromvaf1},
\cite{maldstrom}, \cite{johnmyers}, \cite{bmpv2}, \cite{maldhoro},
\cite{bmpv3}, \cite{maldlowe}. In the case of five-dimensional
black holes, the derivation of the entropy relies
on establishing a correspondence between black holes and string states.
Clearly, this is more straightforward if there is a black hole
uniqueness theorem which constrains the types of possible black hole
solutions. However, such uniqueness cannot be taken for
granted in five (or more) dimensions, and some explicit black ring
solutions have been constructed which violate uniqueness \cite{harvemp},
\cite{empb}; these solutions do not however preserve any supersymmetry.
Although there has been progress in establishing a correspondence between
string states and certain types of black ring solutions \cite{empelv},
in general the lack of black hole uniqueness in higher
dimensions makes the construction of the black hole/string state
correspondence rather more complicated.
It has, however, been possible to
construct a uniqueness theorem for supersymmetric black hole solutions
of the minimal ungauged five-dimensional supergravity \cite{harvey1}.
The theorem classifies all near-horizon geometries
of solutions which have a regular horizon;
the possible regular near-horizon geometries are
flat space, $AdS_3 \times S^2$ or the (under-rotating)
BMPV near-horizon geometry. In addition, the only solution which has
a regular horizon with near-horizon BMPV as its near-horizon geometry
is the BMPV black hole \cite{bmpv2}.

The purpose of this paper is to extend this uniqueness theorem
to include supersymmetric black hole solutions of the ungauged
five-dimensional
supergravity theory coupled to arbitrary many vector multiplets.
This extension is useful because, although the minimal theory
has many interesting properties, it corresponds to a rather
restricted class of higher dimensional solutions. In order to
investigate the higher dimensional physics more fully by compactification
to lower dimensions, one must typically couple the lower dimensional
theory to additional matter. Thus, the extension of the minimal
uniqueness theory to the non-minimal theory under consideration
here represents a  step towards this goal. It is interesting
to note that a (partial) classification of solutions of
gauged five-dimensional supergravity coupled to arbitrary many
vector multiplets constructed in \cite{gutharv1} plays
a crucial role in the extension of the uniqueness theorem.

The plan of this paper is as follows. In section 2.1 we review
some basic properties of the ungauged five-dimensional
supergravity coupled to arbitrary many vector multiplets. The
classification of solutions constructed in \cite{gutharv1} (with
vanishing gauge parameter) is summarized in section 2.2. In
section 2.3 the black hole solutions of this theory constructed in
\cite{sabra1} and \cite{chamsab} are given using the conventions
of section 2.2. The extension of the uniqueness theorem is
presented in section 3. The theorem is split into two parts. In
3.1, the near-horizon geometry of a black hole with a single
regular connected horizon is derived; it is shown that the near-horizon
geometry is locally isometric to flat space, $AdS_3 \times S^2$,
or the near-horizon geometry of the BMPV solution of the minimal theory.
In 3.2 the global properties of the solution with near-horizon BMPV
for its near-horizon geometry are examined, and it is
shown that the only such solution is that found by Chamseddine
and Sabra in \cite{sabra1} and \cite{chamsab}.
Some conclusions are given in section 4.

\sect{Supersymmetric solutions of ${\cal N}=1$ supergravity}

\subsection{${\cal N} = 1$ supergravity}

The action of ${\cal N}=1$ $D=5$ ungauged supergravity coupled to
abelian vector multiplets with scalars taking values in a symmetric space
is
\be
 S = {1 \over 16 \pi G} \int \left( {}^5 R  -Q_{IJ} F^I
 \wedge *F^J -Q_{IJ} dX^I \wedge
 \star dX^J -{1 \over 6} C_{IJK} F^I \wedge F^J \wedge A^K \right)
\ee
where we use a positive signature metric and the fermions have
been set to zero. $I,J,K$ take values $1
\ldots n$ and $C_{IJK}$ are constants that are symmetric on $IJK$ and obey
\be
\label{eqn:jordan}
C_{IJK} C_{J' (LM}
C_{PQ) K'} \delta^{J J'} \delta^{K K'} = {4 \over 3} \delta_{I (L}
C_{MPQ)}.
\ee
The $X^I$ are scalars which are constrained
via
\be
\label{eqn:conda}
{1 \over 6}C_{IJK} X^I X^J X^K=1 \ .
\ee
We may regard the $X^I$ as being functions of $n-1$
unconstrained scalars $\phi^a$. It is convenient to define
\be
X_I \equiv {1 \over 6}C_{IJK} X^J X^K
\ee
so that the condition ({\ref{eqn:conda}}) becomes
\be
X_I X^I =1 \ .
\ee
In addition, the coupling $Q_{IJ}$ depends on the scalars via
\be
Q_{IJ} = {9 \over 2} X_I X_J -{1 \over 2}C_{IJK} X^K \ .
\ee
Additional useful identities which are satisfied
as a consequence of the  Very Special geometry
can be found in \cite{gutharv1}.

The Einstein equation is given by
\bea
\label{eqn:ein}
-{}^5 R_{\alpha \beta} +Q_{IJ} F^I{}_{\alpha \lambda}
F^J{}_\beta{}^\lambda+Q_{IJ} \nabla_\alpha X^I \nabla_\beta X^J
-{1 \over 6}g_{\alpha \beta} \left(Q_{IJ} F^I{}_{\mu \nu}
 F^{J \mu \nu} \right) =0
\eea
the gauge equations are
\be
\label{eqn:gauge}
d \left(Q_{IJ} \star F^J \right)=-{1 \over 4}C_{IJK} F^J \wedge F^K \ ,
\ee
and the scalar equation can be written as
\bea
d \left(\star dX_I \right) - \left({1 \over 6} C_{MNI} -{1 \over
2}X_I C_{MNJ} X^J \right) dX^M \wedge \star dX^N \nn
+ \left( X_M X^P C_{NPI}-{1 \over 6}C_{MNI}-6 X_I X_M X_N+{1 \over 6}
X_I C_{MNJ} X^J \right) F^M \wedge \star F^N=0 .
\eea

In addition, for a bosonic background to be supersymmetric there must be a
spinor $\epsilon^a$. From this Killing spinor we can construct
tensors from spinor bilinears, which can be used to classify the general
supersymmetric solutions of this theory. This classification was
presented in \cite{gutharv1} for the solutions of the gauged theory;
we shall recap the main results which are somewhat simpler in the
case of the ungauged theory. In particular, we obtain a scalar $f$,
a vector $V$ and three 2-forms $J^{(i)}$ which satisfy
the algebraic relations
\be
\label{eqn:Vsq}
 V_{\alpha} V^{\alpha} = -f^2,
\ee
\be \label{eqn:XwedgeX}
 J^{(i)} \wedge J^{(j)} = -2\delta_{ij} f \star V,
\ee \be \label{eqn:VdotX}
 i_V J^{(i)} = 0,
\ee \be \label{eqn:VstarX}
 i_V \star J^{(i)} = - f J^{(i)},
\ee \be \label{eqn:XcontX}
 J^{(i)}_{\gamma \alpha} J^{(j)}_\beta{}^\gamma = \delta_{ij} \left(
 f^2 \eta_{\alpha\beta} + V_{\alpha} V_{\beta} \right) - \epsilon_{ijk} f
 J^{(k)}_{\alpha\beta}
\ee
where $\epsilon_{123} = +1$ and, for a vector $Y$ and $p$-form $A$,
$(i_Y A)_{\alpha_1 \ldots \alpha_{p-1}} \equiv Y^{\beta} A_{\beta \alpha_1
\ldots \alpha_{p-1}}$. In addition to these algebraic relations,
the bilinears also satisfy differential constraints as a consequence of
the gravitino and dilatino equations. These differential
constraints were computed for the more general gauged theory in
\cite{gutharv1}, and the equations for the ungauged theory
can be obtained from those in \cite{gutharv1} by setting the gauge
parameter $\chi$ to vanish; in particular, we find that $V$ is
a Killing vector satisfying
\be
\cL_V f = 0 \ , \quad \cL_V V =0 \ , \quad \cL_V F^I = \cL_V J^i=0 \ .
\ee

\subsection{The timelike case}

It is useful to distinguish
two cases depending on whether the scalar $f$ vanishes everywhere or
not. In the ``null case'', the
vector $V$ is globally a null Killing vector with $f=0$.
As we are interested in
investigating the properties of black hole solutions, we shall
concentrate on the latter ``timelike case'' . Take an open set ${\cal U}$
in which $f$ is positive and hence $V$ is a timelike Killing vector
field.  We shall summarize the constraints imposed by
supersymmetry in the region ${\cal U}$.

Introduce coordinates $(t,x^m)$ such that $V =
\partial/\partial t$. The metric can then be written locally as
\be
 \label{eqn:metric}
 ds^2=-f^2(dt+\omega)^2+f^{-1}h_{mn}dx^m dx^n.
\ee
The metric $h_{mn}$ can be regarded as the metric on a four
dimensional Riemannian manifold, which we shall refer to as the ``base
space" $B$, and $\omega$ is a 1-form on $B$. Since $V$ is Killing,
$f$, $\omega$ and $h$ are independent of $t$. We shall reduce the
necessary and sufficient conditions for supersymmetry to a set of
equations on $B$. Let
\be
 \label{eqn:e0def}
 e^0 = f (dt+\omega).
\ee
We choose the orientation of $B$ so that $e^0 \wedge \eta_4$ is
positively oriented in five dimensions, where $\eta_4$ is the volume
form of $B$. The two form $d\omega$ can be split into self-dual and
anti-self-dual parts on $B$:
\be
\label{eqn:rsp}
f d\omega=G^{+}+G^{-}
\ee
where the factor of $f$ is included for convenience.

Equation ({\ref{eqn:VdotX}}) implies that the $2$-forms $J^{(i)}$ can be
regarded as $2$-forms on the base space and Equation
({\ref{eqn:VstarX}}) implies that they are anti-self-dual:
\be
 \star_4 J^{(i)} = - J^{(i)},
\ee
where $\star_4$ denotes the Hodge dual on $B$.
Equation ({\ref{eqn:XcontX}}) can be written
\be
 \label{eqn:quat}
 J^{(i)}{}_m{}^p J^{(j)}{}_p{}^n = - \delta^{ij} \delta_m{}^n
 + \epsilon_{ijk} J^{(k)}{}_m{}^n
\ee
where indices $m,n, \ldots$ have been raised with $h^{mn}$,
the inverse of $h_{mn}$, so the $J^{(i)}$
satisfy the algebra of imaginary unit quaternions.
In addition, from the differential constraints, we find that
the $J^i$ are covariantly constant on $B$ and so the
base space is hyper-K\"ahler with hyper-complex structures $J^i$.

The differential constraints on the bilinears also  constrain
the gauge field strengths. We find that

\be
\label{eqn:rewritegaug}
F^I = d (X^I e^0) + \Theta^I \ ,
\ee
where $\Theta^I$ is a
self-dual 2-form on $B$ satisfying
\be
\label{eqn:gpluscontr}
X_I \Theta^I = -{2 \over 3} G^+ \ .
\ee

In fact these conditions are sufficient to ensure the existence of
a Killing spinor preserving 4 of the 8 supersymmetries. The Killing
 spinor $\epsilon$ is given by
\be
\label{eqm:killsp}
\epsilon = f^{1 \over 2} \epsilon_0
\ee
where $\epsilon_0$ is covariantly constant on the hyper-K\"ahler
base space and satisfies \footnote{Note that there is
an additional factor of $i$
compared with the expression given in \cite{classa} due to the
change of signature of the metric.}
\be
\label{eqn:newgamma}
\gamma^0 \epsilon = i \epsilon \ .
\ee

However, as we are interested in supersymmetric {\it solutions}
we also need to impose the Bianchi identity $dF^I=0$ and Maxwell
equations ({\ref{eqn:gauge}}).
Substituting the field strengths (\ref{eqn:rewritegaug}) into the
Bianchi identities $dF^I=0$ gives
\be
\label{eqn:bianch}
d  \Theta^I=0 \ ,
\ee
so the $\Theta^I$ are harmonic self-dual 2-forms on the base.
The Maxwell equations ({\ref{eqn:gauge}}) reduce to
\bea
\label{eqn:timegauge}
\nabla^2 \left( f^{-1} X_I \right) &=& {1 \over 6}C_{IJK}
(\Theta^J \ . \ \Theta^K) \ ,
\eea
where $\nabla^2$ denotes the Laplacian on the hyper-K\"ahler base $B$;
and contracting ({\ref{eqn:timegauge}}) with $X^I$ we obtain
\be
\label{eqn:asympthet}
\nabla^2 f^{-1}  =
-{1 \over 3}Q_{IJ} \big( (\Theta^I \ . \  \Theta^J)
+ 2 f^{-1}  (dX^I \ . \ dX^J) \big) +{2 \over 3} (G^+ \ . \ G^+) \ ,
\ee
where we have used the convention that for
p-forms $\alpha$, $\beta$ on $B$, we set
\be
(\alpha \ . \  \beta) = {1 \over p!} \alpha_{m_1 \dots m_p}
\beta^{m_1 \dots m_p}\ .
\ee

The integrability conditions for the existence of a Killing spinor
guarantee that the Einstein equation and scalar equations of motion
are satisfied as a consequence of the above equations.

\subsection{Black Hole Solutions}

Before proceeding with the uniqueness proof, it is useful to recall
the form of the black hole solution found in \cite{sabra1} and
\cite{chamsab}. This solution effectively extends the BMPV
solution of the minimal
ungauged five-dimensional supergravity to include arbitrary
many abelian vector multiplets. In fact, just as for the BMPV solution,
the solution does not only describe a single stationary black hole,
but can be generalized to describe
a multi-centred system of black holes with arbitrary positions.
Here, we shall only consider the single-centre solutions. In our
formalism, the black hole solution is obtained by taking
the base space to be $B= \bR^4$ equipped with metric
\be
ds_4{}^2 = d\rho^2 + {\rho^2 \over 4} \big[ (\sigma^R_1)^2+(\sigma^R_2)^2+
(\sigma^R_3)^2 \big]
\ee
where $\sigma^R_i$ are left-invariant 1-forms on $SU(2)$ given in terms
of the Euler angles $\theta$, $\phi$, $\psi$ by
\bea
\sigma^R_1 &=& - \sin \psi d \theta + \cos \psi \sin \theta d \phi
\nn
\sigma^R_2 &=& \cos \psi d \theta + \sin \psi \sin \theta d \phi
\nn
\sigma^R_3 &=& d \psi + \cos \theta d \phi
\eea
for $0 \leq \theta \leq \pi$, $0 \leq \phi \leq 2 \pi$,
$0 \leq \psi < 4 \pi$. Positive orientation is taken with
respect to ${\rho^3 \over 8}  d\rho \wedge \sigma^R_1 \wedge \sigma^R_2
\wedge \sigma^R_3$. In addition, we take
$\Theta^I=G^+ =0$. Hence the equations ({\ref{eqn:timegauge}}) simplify to
\be
\nabla^2 (f^{-1} X_I)=0
\ee
where $\nabla^2$ denotes the Laplacian on $\bR^4$. These equations
are solved by taking
\be
f^{-1} X_I = \nu_I +{\mu_I \over \rho^2}
\ee
for constants $\mu_I$, $\nu_I$ and hence $f$ is given by
\bea
f^{-3} &=& {9 \over 2} C^{IJK} (1 +{\mu_I \over \rho^2})
(1 +{\mu_J \over \rho^2})(1 +{\mu_K \over \rho^2})
\nn
&=&\alpha_0 +{\alpha_1 \over \rho^2}+{\alpha_2 \over \rho^4}+
{\alpha_3 \over \rho^6}
\eea
where $C^{IJK} = \delta^{II'} \delta^{JJ'} \delta^{KK'} C_{I'J'K'}$ and
\bea
\alpha_0 &=& {9 \over 2} C^{IJK} \nu_I \nu_J \nu_K
\nn
\alpha_1 &=& {27 \over 2} C^{IJK} \mu_I \nu_J \nu_K
\nn
\alpha_2 &=& {27 \over 2} C^{IJK} \mu_I \mu_J \nu_K
\nn
\alpha_3 &=& {9 \over 2} C^{IJK} \mu_I \mu_J \mu_K \ ,
\eea
where we have made use of the identity
\be
X^I={9 \over 2}C^{IJK} X_J X_K \ .
\ee
We require that $f>0$ for $\rho>0$, hence we must take
$\alpha_0 \geq 0$ and $\alpha_3 \geq 0$; and to obtain an
asymptotically flat solution we require $\alpha_0>0$.
By rescaling $t$, we can without loss of generality set $\alpha_0=1$.

Lastly, we require that $d \omega$ be anti-self-dual on $\bR^4$, so we set
\be
\label{eqn:omsol}
\omega = {j \over 2 \rho^2} \sigma^R_3 \ ,
\ee
for constant $j$.
Observe that in order for the closed timelike curves to lie strictly
within the horizon, we require that
$f^{-3} -{j^2 \rho^{-6}}>0$ for $\rho>0$, hence, in particular,
$j^2 < \alpha_3$.

Note that the near-horizon geometry of the above black hole solutions is
given by taking
\be
f^{-1} X_I = {\mu_I \over \rho^2}
\ee
with
\be
f^{-3} = {\alpha_3 \over \rho^6}
\ee
and $\omega$ is given by ({\ref{eqn:omsol}}).

\section{Black Hole Uniqueness}

In order to construct a uniqueness proof, we shall follow the
methodology set out in \cite{harvey1}. In particular, we first
show that the near-horizon geometry of a solution with a regular horizon
is locally isometric to either flat space, $AdS_3 \times S^2$ or
the under-rotating near-horizon BMPV solution. Then, making use of the
fact that the base space is hyper-K\"ahler, the global properties
of the solution which has near-horizon BMPV for its
near-horizon geometry are investigated.

\subsection{The near horizon geometry}

Following the reasoning set out in section 3.3 of \cite{harvey1},
we shall introduce Gaussian null co-ordinates $u,r,x^A$ for $A=1,2,3$
in a neighbourhood of the horizon, so that
\be
V = {\partial \over \partial u}
\ee
and
\be
f = r \Delta (r, x^A)
\ee
with
\be
ds^2 = -r^2 \Delta^2 du^2 +2 du dr +2r h_A du dx^A + \gamma_{AB} dx^A dx^B
\ee
where $h_A=h_A (r, x^A)$, $\gamma_{AB} = \gamma_{AB} (r,x^A)$.
$\Delta$, $h_A$  and $\gamma_{AB}$ are smooth at the horizon;
$\gamma_{AB}$ defines a globally well-defined
metric on a smooth Riemannian 3-manifold in the near-horizon limit, and
$\Delta>0$ for $r>0$. In the minimal theory, these
assumptions were sufficient to ensure that the graviphoton gauge
field strength is regular on the horizon. In the more general theory
which we consider here, we shall see that regularity of the metric on the
horizon is sufficient to prove that $X_I F^I$ is regular on the horizon.
However, in order to construct
the uniqueness proof we shall in fact assume
a stronger condition on the fields, namely that all of the $X^I$ and
all components of the $F^I$ are regular at $r=0$ in the Gaussian null
co-ordinates.

Next, note that the three hyper-K\"ahler forms can be written as
\be
J^i = dr \wedge Z^i + r(h \wedge Z^i- \Delta \star_3 Z^i)
\ee
where $\star_3$ denotes the hodge dual defined
with respect to $\gamma_{AB}$
and $Z^i = Z^i_A dx^A$, $h=h_A dx^A$. As the $J^i$ satisfy the algebra
of the imaginary unit quaternions, it is straightforward to show that
the $Z^i$ define an orthonormal basis on the 3-manifold $H$ with metric
$\gamma_{AB}$. The $J^i$ are closed, so we obtain
\be
\label{eqn:dzi}
{\hat{d}} Z^i = -{1 \over 2} \partial_r (r \Delta) \epsilon_{ijk} Z^j
\wedge Z^k +\partial_r (rh) \wedge Z^i - r \Delta \epsilon_{ijk}
\partial_rZ^j \wedge Z^k +rh \wedge \partial_r Z^i
\ee
and
\be
\label{eqn:dhs}
\star_3 {\hat{d}} h- {\hat{d}} \Delta - \Delta h +r \partial_r \Delta h-
2r \Delta \partial_r h
-r \star_3 (h \wedge \partial_r h)-r \Delta^2 \epsilon_{ijk}Z^i
<Z^j , \partial_r Z^k>=0
\ee
where if $Y$ is a p-form of the type
\be
Y= {1 \over p!} Y_{A_1 \dots A_p} dx^{A_1} \wedge \dots \wedge dx^{A_p}
\ee
we define
\be
{\hat{d}} Y = {1 \over (p+1)!} (p+1) \partial_{[A_1}
Y_{A_2 \dots A_{p+1}]} dx^{A_1} \wedge \dots \wedge dx^{A_{p+1}}\ .
\ee
and $<,>$ denotes the inner product on $H$ with respect to the metric
$\gamma_{AB}$.
In addition, for $r>0$ we obtain
\be
\omega = -{1 \over r^2 \Delta^2} dr -{1 \over r \Delta^2}h
\ee
and hence
\be
\label{eqn:gplusa}
G^+ = dr \wedge {\cal{G}} +r(h \wedge {\cal{G}}+ \Delta \star_3 {\cal{G}})
\ee
where
\be
\label{eqn:gplusb}
{\cal{G}} =-{3 \over 2 r \Delta^2}{\hat{d}} \Delta +{3 \over 2 \Delta^2}
\partial_r \Delta h -{3 \over 2 \Delta} \partial_r h -{1 \over 2}
\epsilon_{ijk} Z^i <Z^j , \partial_r Z^k> \ .
\ee
All of the above identities are identical to those found in \cite{harvey1}
for the minimal theory.
To extend the near-horizon classification to the more general theory,
it is convenient to set
\be
\Theta^I =-{2 \over 3} X^I G^+ + N^I
\ee
so that $X_I N^I =0$. Setting
\be
N^I = dr \wedge T^I +r(h \wedge T^I + \Delta \star_3 T^I)
\ee
for $T^I=T^I_A dx^A$, and using ({\ref{eqn:rewritegaug}}) we find

\bea
\label{eqn:fieq}
F^I &=& \big[\partial_r (r \Delta X^I) dr + r {\hat{d}} (\Delta X^I) \big]
\wedge du + dr \wedge Q^I + rh \wedge Q^I
\nn
&+& r \Delta \star_3 T^I -X^I \star_3 h -r X^I \star_3 \partial_r h
-{2 \over 3} \Delta r \epsilon_{ijk} \star_3 Z^i <Z^j , \partial_r Z^k>
\eea
where
\be
Q^I =  T^I +{1 \over r \Delta} {\hat{d}} X^I
-{1 \over \Delta} \partial_r X^I h
+{1 \over 3} X^I \epsilon_{ijk} Z^i <Z^j , \partial_r Z^k>
\ee
On using $X_I T^I=0$, $X_I {\hat{d}} X^I=0$ and $X_I \partial_r X^I =0$
we find that $X_I F^I$ is regular on the horizon. However, our assumption
of the regularity of both $X^I$ and $F^I$ is somewhat stronger. In
particular we find  that $Q^I$ is regular on the horizon.
Note that the spatial components of $F^I$ are given by
\bea
{1 \over 2} F^I_{AB} dx^A \wedge dx^B &=& r \Delta \star_3 Q^I -
\star_3 {\hat{d}} X^I +r \partial_r X^I \star_3 h -X^I \star_3 h
\nn
&-&r X^I \star_3 \partial_r h -{2 \over 3} \Delta r \epsilon_{ijk}
\star_3 Z^i <Z^j , \partial_r Z^k>
\eea
Hence, evaluating the spatial components of the Bianchi identity at
$r=0$ we find
\be
{\hat{d}} \big(X^I \star_3 h + \star_3 {\hat{d}} X^I \big)=O(r)
\ee
Contracting this expression with $X_I$ we obtain
\be
{\hat{d}} \star_3 h +{2 \over 3} Q_{IJ} {\hat{d}} X^I \wedge \star_3
{\hat{d}} X^J = O(r) \ .
\ee
Integrating over $H$ we find that
\be
\int_{H} Q_{IJ} {\hat{d}} X^I \wedge \star_3 {\hat{d}} X^J = O(r)
\ee
and as $Q_{IJ}$ defines a positive-definite inner product it follows that
\be
{\hat{d}} X^I = O(r)
\ee
and hence
\be
{\hat{d}} \star_3 h = O(r) \ .
\ee

So, it follows that $h$, $\Delta$ and the $Z^i$ satisfy
exactly the same constraints on the horizon as in the minimal theory.
These constraints were analysed in detail in \cite{harvey1}
so we shall only present the results of that analysis here.

In particular, it is straightforward to see that $\Delta$ must be
constant on the horizon. The case for which $\Delta=0$ on the
horizon is special. There are then two sub-cases; in the first $h
\neq 0$ on the horizon and the near-horizon geometry is locally
isometric to $AdS_3 \times S^2$. It was originally argued in the
appendix of \cite{harvey1} that solutions with this near-horizon
geometry can be excluded; however, more recent work in
\cite{blackring} has shown that there indeed exist asymptotically
flat supersymmetric black ring solutions with $AdS_3 \times S^2$
near-horizon geometry. In the second sub-case, $h=0$ on the horizon
and the near-horizon geometry is locally isometric to Minkowski space;
it has not yet been determined whether this geometry can arise as
the near-horizon geometry of a black hole.

Notwithstanding this difficulty, we shall assume henceforth that
$\Delta>0$ on the horizon.
Then from section 3.6 of \cite{harvey1} it is straightforward to
see that local co-ordinates $\phi$, $\theta$, $\psi'$ can be
introduced with respect to which the metric on $H$ can be written as
\be
ds_3^2 = {\mu \over 4} \big[ (1-{j^2 \over \mu^3})
(d \psi' + \cos \theta d \phi)^2
+d \theta^2+ \sin^2 \theta d \phi^2 \big]
\ee
with
\be
h = -j \mu^{-{3 \over 2}} \sqrt{1-{j^2 \over \mu^3}}
(d \psi' + \cos \theta d \phi)
\ee
where $\mu$ and $j$ are constants with $\mu>0$, $j^2 < \mu^3$ and
\be
\Delta = 2 \mu^{-{1 \over 2}} \sqrt{1-{j^2 \over \mu^3}} \ .
\ee

To summarize, we have shown that in the near-horizon limit, $X^I$
and $\Delta$ are constant. Assuming that $\Delta \neq 0$ on the horizon,
we have proven that the near-horizon geometry
is locally isometric to that of the near-horizon BMPV solution
in the minimal theory. This is also the near-horizon geometry of the
Chamseddine and Sabra black holes. In particular, if $j=0$ then $H$
is locally isometric to the round 3-sphere. Globally, we must have
$H= S^3/\Gamma$ where $\Gamma$ is a discrete subgroup of $SU(2)_R$.
If $j \neq 0$ then H is locally isometric to a squashed 3-sphere;
globally $H=S^3/ \Gamma$ where $\Gamma$ is a cyclic group.

\subsection{Global Analysis}

Given the similarity between the analysis of the near-horizon geometries
in the general and the minimal five-dimensional supergravity theories,
it is unsurprising that there is also a considerable similarity in the
global analysis of the black hole solutions for which the near-horizon
geometry is the near-horizon BMPV geometry.
In particular, if we make a change of co-ordinates
\bea
R &=& (2r)^{1 \over 2} \mu^{1 \over 4} (1-{j^2 \over \mu^3})^{1 \over 4}
\nn
\psi &=& \psi' -2j \mu^{-{3 \over 2}} (1-{j^2 \over \mu^3})^{-{1 \over 2}}
\log R
\eea
then locally, in a neighbourhood of the horizon, the metric on the
hyper-K\"ahler base takes the form
\bea
ds_4^2 &=& dR^2 +{R^2 \over 4}\big[ (d \psi+\cos \theta d \phi)^2+
d \theta^2+ \sin^2 \theta d \phi^2 \big]
\nn
&+& O(R^2) dR^2 + O(R^3) dR dy^A +O(R^4) dy^A dy^B
\eea
where $y^A = (\psi , \theta , \phi)$. Hence, locally, the base space
is flat. However, it is, in principle,
possible for there to be a conical singularity at $R=0$. In fact,
just as in the case of the minimal theory \cite{harvey1},
the conical singularity  must be an $A-D-E$ orbifold singularity
because the base is hyper-K\"ahler.
Moreover, we also require that the solution be asymptotically
flat, i.e. $f$ must tend to a positive constant at infinity,
and $\omega$ must decay sufficiently rapidly so that
the ADM angular momentum is well-defined. Hence
the base space must be asymptotically Euclidean.
Now, the $A-D-E$ orbifold singularities at $R=0$ can
be resolved \cite{aspwall} by blowing up the singularity.
Thus we obtain a new hyper-K\"ahler base space which
is complete and asymptotically Euclidean. It is known that
only one such manifold exists \cite{gibpoper4};
the base must be globally $\bR^4$ equipped with metric
\be
ds_4^2 = d \rho^2 + {\rho^2 \over 4} \big[
(d \psi+\cos \theta d \phi)^2+ d \theta^2+ \sin^2 \theta d \phi^2 \big]
\ee
and
\be
R = \rho +O(\rho^3)
\ee
where $0 \leq \theta \leq \pi$ and $\phi \sim \phi+2 \pi$,
$\psi \sim \psi + 4 \pi$.

The next step in the global analysis is to show that
$\Theta^I=0$. To see this, note first that
\be
F^I = d(r \Delta X^I) \wedge du -d ({X^I \over r \Delta}) \wedge dr
 -d({X^I \over \Delta} h)+ \Theta^I \ .
\ee
Moreover, we recall that as $X^I$ and $\Delta$ are constant on
the horizon (with $\Delta \neq 0$ on the horizon), it follows that
the $d ({X^I \over r \Delta}) \wedge dr$ term in this
expression is regular on the horizon. Hence, as we assume that
$F^I$ is regular at the horizon,
 $\Theta^I$ (and hence also $G^+$) is regular at the
origin of the base space. Moreover,
$G^+$ must vanish at infinity in $\bR^4$ due to the asymptotic
decay of $\omega$. In addition, $f$ tends to a positive constant at
infinity with a decay sufficient to ensure the existence of a
well-defined ADM mass. Using these facts, it is
straightforward to see from ({\ref{eqn:asympthet}}) that
$Q_{IJ} \big( (\Theta^I \ . \ \Theta^J)
+ 2 f^{-1}  (dX^I \ . \ dX^J) \big)$
must vanish at infinity. Hence, at infinity the $\Theta^I$ must vanish
and the $X^I$ are constant.
As $\Theta^I$ is closed and globally defined on $\bR^4$
it follows that there exists a 1-form
$\Lambda^I$ also globally defined on $\bR^4$,
 with $\Lambda^I$ vanishing at infinity,
such that $\Theta^I = d \Lambda^I$. Then
\be
0 = \int_{S^3_\infty} \Lambda^I \wedge \Theta^I =
\int_{\bR^4} \Theta^I \wedge \Theta^I = \int_{\bR^4}
(\Theta^I \ . \ \Theta^I)
\ee
where $S^3_\infty$ denotes the 3-sphere at infinity in $\bR^4$.
Hence it follows that $\Theta^I=G^+=0$.

As $\Theta^I=0$, we find from ({\ref{eqn:timegauge}})
that the $f^{-1} X_I$ are harmonic on $\bR^4$. Suppose that
$X_I \rightarrow \mu^{-1} \mu_I $ as
$R \rightarrow 0$ and $X_I \rightarrow \nu_I$ as
$R \rightarrow \infty$ for constants $\mu_I$,
$\nu_I$. Then in a neighbourhood of the origin we
have $f \sim {\rho^2 \over \mu}$, and hence
\be
\label{eqn:fexpr}
f^{-1} X_I = {\mu_I \over \rho^2} + g_I
\ee
where $g_I$ is $O(\rho^0)$ near $\rho=0$. As
$f^{-1} X_I$ is regular outside the horizon, we
note that $g_I$ must be regular, bounded and harmonic
on $\bR^4$. Hence $g_I$ is constant.
By re-scaling $u$ we can without loss of generality set
$f \rightarrow 1$ as $R \rightarrow \infty$, so we must have $g_I= \nu_I$.
Observe that ({\ref{eqn:fexpr}}) implies that
\bea
f^{-3} &=& {9 \over 2} C^{IJK} (1 +{\mu_I \over \rho^2})
(1 +{\mu_J \over \rho^2})(1 +{\mu_K \over \rho^2})
\nn
&=&\alpha_0 +{\alpha_1 \over \rho^2}+{\alpha_2 \over \rho^4}
+ {\alpha_3 \over \rho^6}
\eea
where
\bea
\alpha_0 &=& {9 \over 2} C^{IJK} \nu_I \nu_J \nu_K
\nn
\alpha_1 &=& {27 \over 2} C^{IJK} \mu_I \nu_J \nu_K
\nn
\alpha_2 &=& {27 \over 2} C^{IJK} \mu_I \mu_J \nu_K
\nn
\alpha_3 &=& {9 \over 2} C^{IJK} \mu_I \mu_J \mu_K \ .
\eea
Hence we must have $\alpha_0=1$ and $\alpha_3 = \mu^3$.
Observe that $j^2 < \alpha_3$.

Finally, we note that by exactly the same reasoning
as set out in section 3.6 of
\cite{harvey1}, we must have
\be
\omega = {j \over 2 \rho^2} \sigma^R_3 + db
\ee
for some function $b$. As $b$ can be absorbed into
the time co-ordinate $t$ by a shift, we can set
without loss of generality
\be
\omega = {j \over 2 \rho^2} \sigma^R_3 \ .
\ee

Hence we have proven that the only regular black hole solution
of this theory with a near-horizon geometry which is
the near-horizon BMPV geometry (in which limit the
$F^I$ and $X^I$ are sufficiently regular) is
the under-rotating (i.e. $j^2 < \mu^3$) black hole solution
of Chamseddine and Sabra \cite{sabra1}, \cite{chamsab}.

\section{Conclusions}

In this paper we have presented an extension of the black
hole uniqueness theorem of \cite{harvey1} to a more general
ungauged five-dimensional supergravity coupled to arbitrary many
abelian vector multiplets. It is remarkable that the Very Special
geometry associated with this theory imposes sufficiently
strong constraints on supersymmetric solutions to allow
for such a theorem to be proved. The only sufficiently regular
solutions of the theory which have an event horizon
locally isometric to the near-horizon BMPV geometry
are the black holes found in
\cite{sabra1} and \cite{chamsab}.
It would be interesting to determine whether or not
the near-horizon solutions with $\Delta=0$ and $h=0$ on the
horizon can arise as the near-horizon geometry of an
asymptotically flat supersymmetric black hole.
Also, in order to prove the theorem,
rather strong regularity conditions were imposed on the scalars and
field strengths at the horizon. It might be possible to weaken
these conditions and investigate whether additional black hole
solutions exist.

Another interesting problem would be to attempt to
prove a black hole uniqueness theorem in gauged five-dimensional
supergravity. Recently, the first examples of
regular supersymmetric black holes
of such a theory have been found in \cite{gutharv1} and \cite{gutharv2}.
However, even for the case of the minimal gauged theory, attempting to
adapt the methods used in the ungauged uniqueness theorem is
somewhat problematic. In particular, for the near-horizon analysis, it
is by no means apparent that the scalar $\Delta$ must be constant on the
horizon. More seriously, the base space in the gauged theory is not
hyper-K\"ahler, rather it is only K\"ahler. Hence the global analysis
which was used for the ungauged theory cannot be straightforwardly
generalized. Finally, we remark
that there are more general five-dimensional
supergravity theories; a recent useful discussion of this can be
found in \cite{genfive}. It would be useful to determine if these
more general theories admit new black hole solutions.

\acknowledgments{J. G. was supported by EPSRC. I thank Harvey Reall
for useful discussions.}

\end{document}